\documentclass[referee]{raa}            

\usepackage{graphicx,times}             
\usepackage{natbib}
\usepackage{amssymb,amsmath}
\bibpunct{(}{)}{;}{a}{}{,}

\usepackage[pagebackref=true]{hyperref}
\usepackage{orcidlink}
\usepackage{booktabs}
\usepackage{graphicx}
\usepackage{xcolor}
\usepackage{array}
\usepackage{amsmath}
\usepackage{tabularx}  
\usepackage{float}
\newcommand{\rev}[1]{\textcolor{black}{#1}}

\newcommand{\xycao}[1]{\textcolor{black}{#1}}

\begin{document}

  \title{CSST Strong Lensing Preparation: Cosmological constraints from double-source-plane strong lensing systems in era of CSST}

   \volnopage{Vol.0 (20xx) No.0, 000--000}      
   \setcounter{page}{1}          

   \author{Bei-chen Wu 
      \inst{1,2,3}
   \and Xiaoyue Cao
      \inst{4}
   \and Nan Li \orcidlink{0000-0001-6800-7389}
      \inst{1,2,3}
   \and Yan Gong \orcidlink{0000-0003-0709-0101}
      \inst{1,2,3}
   \and Shenzhe Cui
      \inst{1,2,3}
   \and Di Wu
      \inst{1,2,3}
   \and Tong Zhao
      \inst{1,2}
   \and Junhui Yan
      \inst{1,2,3}
   }

   \institute{National Astronomical Observatories, Chinese Academy of Sciences, Beijing 100101, People's Republic of China; {\it nan.li@nao.cas.cn; gongyan@bao.ac.cn}\\
        \and
        University of Chinese Academy of Sciences, Beijing 100049, People's Republic of China\\
        \and
        Science Center for China Space Station Telescope, National Astronomical Observatories, Chinese Academy of Sciences, 20A Datun Road, Beijing 100101, China\\
        \and
        Institute for Astrophysics, School of Physics, Zhengzhou University, Zhengzhou, 450001, China
\vs\no
   {\small Received 20xx month day; accepted 20xx month day}}

\abstract{ 
\xycao{Double source plane strong lensing (DSPL) systems offer a robust, independent probe of cosmological parameters. The Chinese Space Station Telescope (CSST) is expected to discover hundreds of DSPLs, yet the survey modes and system configurations that best enable cosmological inference remain uncertain. To investigate the impact of varying signal-to-noise ratios (SNR) and Einstein radius ratios of DSPLs (denoted as $\beta^{-1}$ parameters) on cosmographic inference under different CSST survey modes (Wide Field (WF), Deep Field (DF), and Ultra-Deep Field (UDF)), we simulate and model mock lenses with Singular Isothermal Ellipsoid (SIE) mass profiles and S\'ersic sources whose image properties are tailored to CSST specifications. Assuming a flat $w$CDM universe with fiducial values $\Omega_{\rm m} = 0.30966$ and $w = -1$, and uniform priors of $\Omega_{\rm m} \in [0, 1]$ and $w \in [-2, -1/3$), we find that the constraining power on cosmological parameters for a given DSPL system increases significantly with survey depth. For a representative DSPL system with two prominent arcs and a moderate $\beta^{-1}=1.17$, the constraints on ($w, \Omega_{\rm m}$) improve from ($-1.28_{-1.00}^{+0.64}, 0.50_{-0.32}^{+0.28}$) in the WF to ($-1.59_{-0.32}^{+0.63}, 0.42_{-0.06}^{+0.15}$) in the UDF. Furthermore, we find that systems with smaller $\beta$ values yield tighter cosmographic constraints. We conclude that DSPL systems identified in UDF observations, particularly those with small $\beta$, are the most promising candidates for early-stage cosmological studies with CSST.}
 }

   \authorrunning{B.-C. Wu, X.-Y. Cao \& N. Li et al.}            
   \titlerunning{Cosmological constraints from DSPL in CSST era}  

   \maketitle
%
%
\section{Introduction}
\label{sect:intro}

Strong gravitational lensing is a powerful tool for investigating the nature of dark matter and dark energy as well as the formation and evolution of galaxies and galaxy clusters. For instance, it can be utilized to constrain the mass distribution of galaxies and galaxy clusters \citep{Taylor_1998, Treu_2004, clusters_2024}, search for dark matter substructures \citep{he_2022, vegetti_2012}, reveal the properties of high-redshift galaxies \xycao{\citep{Stacey2025, Rizzo2020, Cheng2020, hezaveh_2013, Newton2011}}, and perform statistical detections and constrain cosmological parameters \citep{oguri_2007, asantha_1998, collett_2014}, among various other applications.

In the context of using strong gravitational lensing systems to constrain cosmological parameters, a Double Source Plane Lensing system (DSPLs), with two background sources at different redshifts, serves as a valuable and complementary probe. Unlike time-delay cosmography which scales with the Hubble constant, the ratio of angular separations between images from the first source and the second source, denoted as $\beta$, acts as a geometrical probe free from uncertainties related to the growth history of cosmic structures, allowing constraints on the dark energy equation of state independent of the Hubble constant \citep{collett_2012}. DSPLs exhibit several advantageous features, notably the ratio $\beta$ involves the distances between the sources and the lens, allowing us to probe the distant universe without dependence on the distance ladder of the local universe. Even at high redshifts, up to $z=5$ or beyond, DSPLs remain an effective probe \citep{sharma_2022}. A more compelling feature of DSPLs is their complementarity with conventional distance measurements, which offers constraints with degeneracy being orthogonal to those of other probes \citep{linder_2016}.

Following the discovery of the first DSPL, the ``Jackpot'' system, by the SLACS project \citep{Bolton2008}, DSPLs have been proposed as powerful probes for cosmography. The use of strong lensing probes is typically compromised by the mass-sheet degeneracy (MSD), or the more general source-position transformation (SPT), which limits their accuracy in constraining cosmological models \citep{Cao2022, Teodori_2022, yild_2023, du_2023}. In DSPLs, two independent sets of lensed images arise from sources at different redshifts; this configuration can partially break these degeneracies and thereby reduce associated systematic uncertainties \citep{collett_2012, schneider_2014}. The mitigation is particularly effective when DSPL constraints are combined with complementary information, such as stellar kinematics (e.g.\ velocity dispersion) or time-delay measurements \citep{dux_2024}. However, only a handful of DSPL systems are known \citep{gavazzi_2008, sahu_2025, Bowden_2025, dux_2024, tanaka_2016, Li_2025}. As a result, current analyses of DSPLs remain limited to individual case studies. Research efforts have primarily focused on refining and jointly modeling the ``Jackpot'' case, incorporating spectroscopic and dynamical information to constrain cosmology \citep{turner_2024, ballard_2023, smith_2021, collett_2020, collett_2014}. Other notable cases include the first Einstein zig-zag lens \citep{dux_2024} and the ``Eye of Horus'' system \citep{tanaka_2020, tanaka_2016}. Most recently, DSPLs from the AGEL Survey have been utilized for cosmology \citep{sahu_2025, Bowden_2025}. Additionally, Euclid (launched in 2023) has reported hundreds of new strong lens candidates in its Quick Release 1 (Q1) data \citep{pearce-casey_2024, gabarra_2024}, including four new DSPL candidates \citep{Li_2025}. 

One of the most significant limitations currently hindering the field is the small sample size of DSPLs, which has prevented them from achieving the precision of other established cosmological probes. Fortunately, the advent of next-generation wide-field surveys, such as the CSST, Euclid, Roman, and LSST, is expected to discover hundreds of thousands of strong lenses \citep{gavazzi_2008, collett_2015, cao_2024}, which potentially contain hundreds to thousands of high-quality DSPL systems \citep{dore_2019, sharma_2023, cao_2024, Li_2025}. In particular, the Chinese Space Station Survey Telescope (CSST) \citep{Zhan_2021, gong1_2025}, a 2-meter telescope scheduled to be launched in the late 2020s \citep{gong2_2025} with an image resolution comparable to the \textit{Hubble Space Telescope}, is expected to detect over 160,000 galaxy-galaxy strong lensing systems and over two hundred of high-quality DSPL systems through its survey of around 17,500 square degrees of the sky \citep{cao_2024}. It will also provide photometric redshifts for both lens and source galaxies, offering a less accurate but more cost-effective alternative to spectroscopic redshifts \citep{gong_2019, cao_2024}. Furthermore, while potentially costly, ground-based follow-up high-precision spectroscopic redshift observations like the 4MOST strong lensing spectroscopic survey can further ensure the capability of DSPLs to constrain cosmology \citep{collett_2023, linder_2015}. The availability of such large samples will allow DSPLs to transition from isolated case studies to robust statistical probes, potentially bringing revolutionary changes to cosmological constraints. 

The level of precision achievable for cosmological constraints using survey data products from missions such as CSST and \emph{Euclid} remains uncertain. Quantitative forecasts are therefore required to assess the cosmological applications of DSPLs in the era of next-generation surveys. In particular, it is not yet clear which CSST survey mode, or which classes of DSPL systems, will be most suitable for cosmological inference. We therefore investigate how the signal-to-noise ratio (SNR) in the CSST wide-field (WF), deep-field (DF), and ultra-deep-field (UDF) survey modes, together with the ratio of the two Einstein radii, affects the precision of DSPL-based cosmological constraints. \rev{Our results will offer guidance for constraining cosmology with DSPLs from CSST, particularly regarding the selection of optimal systems.} This work provides a basis for future discussions on how DSPLs can be exploited by CSST and other major facilities under various observing conditions and within different DSPL-based inference frameworks \citep{gong_2019}.

This paper is organized as follows. We introduce cosmography with DSPL in Section \ref{sec:Theo}. We present the mock image simulation and the modeling methodology in Section \ref{sec:Simulation}; the cosmological constraints under different observational scenarios are analyzed and discussed in Section \ref{sec:results}, followed by discussion and conclusions in Section \ref{sec:conclusion}.

\section{Cosmography with DSPL}
\label{sec:Theo}

\begin{figure}
  \centering
  \includegraphics[width = \linewidth]{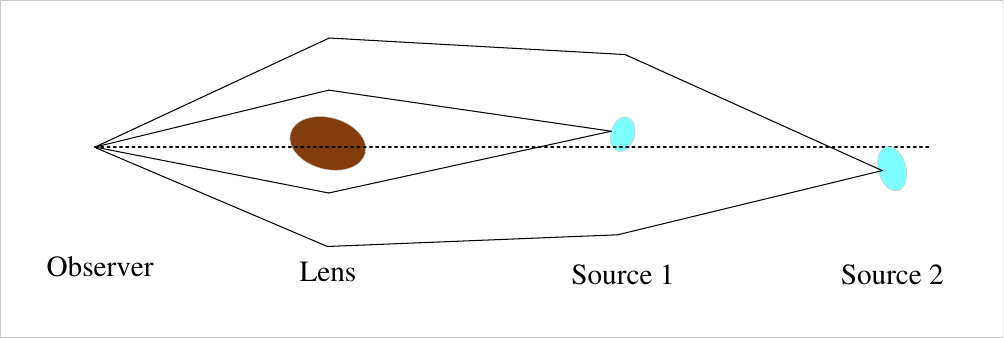}
  \caption{The sketch of a double source plane lensing system (DSPL).}
  \label{fig_s}
\end{figure}

For a double source-plane lens system (see Fig. \ref{fig_s}), the lens equations for the two sources are \citep{collett_2014}:
\begin{equation}\label{eq1}
\mathbf{y}^{\mathrm{S}1}=\mathbf{x}-\beta\,\boldsymbol{\alpha}_1(\mathbf{x}),
\end{equation}
\begin{equation}\label{eq2}
\mathbf{y}^{\mathrm{S}2}=\mathbf{x}-\boldsymbol{\alpha}_1(\mathbf{x})-\boldsymbol{\alpha}_{\mathrm{S}1}\!\left(\mathbf{x}-\beta\,\boldsymbol{\alpha}_1(\mathbf{x})\right),
\end{equation}
where $\mathbf{y}^{\mathrm{S}1}$ and $\mathbf{y}^{\mathrm{S}2}$ are the angular positions on the first (nearer) and second (farther) source planes, respectively, and $\mathbf{x}$ is the angular position in the image plane. Here $\boldsymbol{\alpha}_1(\mathbf{x})$ is the scaled deflection angle produced by the foreground lens, and $\boldsymbol{\alpha}_{\mathrm{S}1}$ is the deflection angle associated with the first source plane, evaluated at the mapped position $\mathbf{x}-\beta\,\boldsymbol{\alpha}_1(\mathbf{x})$. The parameter $\beta$ is the cosmological scaling factor.

For two photons traveling through the same
point in the foreground lens plane, but coming from each source plane, we can derive:
\begin{equation}\label{eq3}
  \begin{aligned}
  \beta\left(z_l, z_{s1}, z_{s2}\right) =\frac{\mathbf{\alpha}(\mathbf{x})}{\mathbf{\alpha_1}(\mathbf{x})} =\frac{D_{l s}\left(z_l, z_{s1}\right)}{D_s\left(z_{s1}\right)} \frac{D_s\left(z_{s2}\right)}{D_{l s}\left(z_l, z_{s2}\right)},
  \end{aligned}
\end{equation}
where the lens is at redshift $z_l$, the first source is located at redshift $z_{s1}$, and the second source at redshift $z_{s2}$. $\mathbf{\alpha}(\mathbf{x})$ is the scaled deflection caused by the foreground lens on the first source plane. $D_s(z)$ denotes the angular diameter distance from the observer to the source at redshift $z$, and $D_{ls}(z_l, z_s)$ denotes the angular diameter distance between the lens and the source.

The cosmological scaling factor $\beta$, hence is determined only by the ratio of angular diameter distances. The foreground lensing effect cancels out in Eq.~\eqref{eq3},  mitigating the degeneracies associated with the foreground lens mass, which is the primary systematic in DSPL. Therefore, reconstructed from data, $\beta$ can serve as a vital and reliable bridge to constrain cosmological parameters. In a simple case where we assume the lens galaxy has a
singular isothermal sphere (SIS) profile, the parameter $\beta$ is the ratio of the two Einstein radii \citep{collett_2012}.

Assuming a flat $w$CDM cosmology, the angular diameter distance is given by:
\rev{
\begin{equation}
  D\left(z_i, z_j\right)=\frac{c / H_0}{\left(1+z_j\right)}\int_{z_i}^{z_j} \mathrm{~d} z^{\prime}\left[\Omega_{\mathrm{m}}\left(1+z^{\prime}\right)^3+\Omega_{\Lambda}(1+z^{\prime})^{3(1+w)}\right]^{-1 / 2},
\end{equation}}
while the Hubble constant ${H_0}$ cancels out in the ratio of angular diameter distances, $\beta$ is able to constrain the dark energy equation of state parameter $w$ and the matter density parameter $\Omega_{\mathrm{m}}$ with the information of redshifts.

The constraining power of DSPLs for cosmological inference depends on the precision with which $\beta$ can be measured. This precision is limited by the signal-to-noise ratio and spatial resolution of the imaging data, as well as by the accuracy of the redshift measurements.

\section{Experiments}
\label{sec:Simulation}
Given the aims of this work, we adopt simplified yet standard lensing models to simulate representative DSPL systems. We neglect sources of systematic uncertainty that are not central to our analysis, including those arising from modelling assumptions, intrinsic degeneracies in the lens and source parametrisations, \rev{contamination from lens light}, environmental contributions, and line-of-sight structures.

To generate mock images for forecasting cosmological constraints under different survey strategies and for DSPL systems with distinct double-ring configurations, we proceed as follows. We model the lens mass distribution with an analytical singular isothermal ellipsoid (SIE) profile and the source surface brightness with a S\'ersic profile. To isolate the effect of the ratio of the two Einstein radii, we select three representative systems from the CSST DSPL sample simulated by \citet{cao_2024}, and fix the remaining lensing parameters accordingly; these systems span a range of Einstein-radius ratios. For each selected system, we then generate mock images for each CSST survey mode by convolving with the appropriate point spread function (PSF) and adding noise according to the corresponding noise model. Across survey modes, the signal-to-noise ratio (SNR) is treated as the only varying parameter.

After generating the mock images, we assume a flat $w$CDM cosmology and apply parametric lens modelling. We then use \texttt{emcee} to infer cosmological parameters from each mock image.

\subsection{Lensing Models}
Our basic model configuration is as follows: The foreground lens is represented by a SIE lens model \citep{kormann_1994}; the light of the first source (source 1) is simulated with a \xycao{S\'ersic} profile \citep{sersic_1963}, and the lensing effect of source 1 to source 2 is modeled using an additional SIE lens model (generally with significantly lower mass compared with the foreground lens) centered at the same coordinates as the source 1 light profile; the light of the second source (source 2) is also represented with a S\'ersic profile.
To ensure our analysis is representative, we adopt the latest cosmological parameters from \cite{planck_2020}, with $\Omega_{\rm m} = 0.30966$, $w = -1$.

As derived by \cite{keeton_2002}, the horizontal and vertical reduced deflection angles $\left(\alpha_1, \alpha_2\right)$ of the \rev{SIE profile, as functions of positions on the lens plane $\left(x_1, x_2\right)$, are given by:
\begin{equation}\label{eq5}
  \begin{aligned}
    &\alpha_1\left(x_1, x_2\right)=b\sqrt{\frac{q}{1-q^2}} \tan ^{-1} \frac{\sqrt{1-q^2} x_1}{\psi},\\
  \end{aligned}
\end{equation}
\begin{equation}\label{eq6}
  \begin{aligned}
    &\alpha_2\left(x_1, x_2\right)=b\sqrt{\frac{q}{1-q^2}} \tanh ^{-1} \frac{\sqrt{1-q^2} x_2}{\psi},
  \end{aligned}
\end{equation}
where $\psi^2=q^2x_1^2+x_2^2$}, and $q$ is the ratio of semi-minor to semi-major axes. \xycao{The normalisation
\begin{equation}
b = 4\pi\left(\frac{\sigma_v}{c}\right)^2\frac{D_{ls}}{D_s}
\end{equation}
is the angular Einstein radius, where $\sigma_v$ is the velocity dispersion of the lens galaxy, $c$ is the speed of light, and $D_{ls}$, $D_s$ represent the angular diameter distances from the lens to the source and from the observer to the source, respectively. }

The S$\acute{\rm e}$rsic profile is given by:
\begin{equation}\label{eq7}
  \begin{aligned}
    I(R)=I_e \exp \left\{-b_n\left[\left(\frac{R}{R_e}\right)^{1 / n}-1\right]\right\},
  \end{aligned}
\end{equation}
where $R_e$ is the half-light radius, and $I_e$ is the intensity at that radius, $n$ is the S$\acute{\rm e}$rsic index and $b_n$ is a dimensionless constant fully determined by $n$ \citep{graham_2005}.

\subsection{Lensing Simulations}
\begin{table}[H]
  \centering
  \setlength{\arrayrulewidth}{0.5mm}
  \setlength{\tabcolsep}{8pt}
  \renewcommand{\arraystretch}{1.4} 
  
  \begin{tabular}{@{}>{\centering\arraybackslash}m{4cm} >{\centering\arraybackslash}m{3cm} >{\centering\arraybackslash}m{3cm} >{\centering\arraybackslash}m{3cm}@{}}
  \toprule
  \textbf{Parameters} & \textbf{System 1 ($\beta^{-1}$ = 1.92)} & \textbf{System 2 ($\beta^{-1}$ = 1.17)} & \textbf{System 3 ($\beta^{-1}$ = 1.02)} \\
  \midrule
  \multicolumn{4}{c}{\textbf{Input Lens: SIE}} \\
  \midrule
  Center $(x, y)$ [($''$, $''$)]  & (0.17, -0.14) & (-0.11, 0.04) & (-0.07, -0.01) \\
  $\theta_E$ [$''$]     &1.66 & 1.29 & 1.53 \\
  Axis ratio             &0.92 &0.88 &0.84 \\ 
  Position angle [$^\circ$] & 83  &-56  &13 \\
  Redshift               &0.33 &0.56  &0.54 \\
  \midrule
  \multicolumn{4}{c}{\textbf{Input Source 1: SIE + S$\acute{\rm e}$rsic}} \\
  \midrule
  Center $(x, y)$ [($''$, $''$)]  & (-0.26, -0.34) & (0.08, 0.27) & (-0.23, -0.04) \\
  $\theta_E$ [$''$]     & 0.032 & 0.008 & 0.004 \\
  Axis ratio             &0.58 &0.69 &0.42 \\
  Position angle [$^\circ$] &71 &-25 &-9 \\
  $I'$ [$e^{-} \, \text{pix}^{-1} \, \text{s}^{-1}$] &14.9 &5.9 &3.3 \\
  \rev{$R_e$} [$''$]           &0.24 &0.14 &0.13 \\
  $n$                    &1.07 &1.20 &1.16 \\
  Redshift               &0.54 &0.99 &2.67 \\
  \midrule
  \multicolumn{4}{c}{\textbf{Input Source 2: S$\acute{\rm e}$rsic}} \\
  \midrule
  Center $(x, y)$ [($''$, $''$)]  &(0.17, -0.07) &(0.05, -0.35) &(0.65, -0.69) \\
  $I'$ [$e^{-} \, \text{pix}^{-1} \, \text{s}^{-1}$] &16.0 &22.4 &14.9 \\
  \rev{$R_e$} [$''$]           &0.05 &0.07 &0.18 \\
  $n$                    &1.04 &1.10 &1.08 \\
  Axis ratio             &0.78 &0.73 &0.70 \\
  Position angle [$^\circ$] &35 &12 &-70 \\
  Redshift               &1.25 &1.14 &2.97 \\
  \bottomrule
  \end{tabular}
  \caption{Lensing simulation parameters for high-, medium-, and low-quality mock systems, with $\beta^{-1}$ values ranked from highest to lowest.}
  \label{Tab0}
\end{table}
To investigate the impact of different ratios of the two Einstein radii on cosmological constraints, based on the images and the value of $\beta$, we selected three typical systems from the sample simulated by \cite{cao_2024}. To ensure representativeness, three systems are selected based on their $\beta^{-1}$ values: one from the top 10\% (high-quality system), one from the 45th--55th percentiles (medium-quality system), and one from the bottom 10\% (low-quality system). The resulting parameters are shown in Table~\ref{Tab0}. The mass and light profiles of source 1 have the same center ($x, y$) coordinates. Note that each value is rounded to 1-3 significant figures.


For the selected systems, we simulate images on a $100\times100$ grid with a pixel scale of $0.074\,\arcsec$ for the CSST main survey. To compute the flux in each pixel accurately, we evaluate the model on an oversampled $16\times16$ subgrid within each native pixel and then average the subpixel values to obtain the native image pixel value. The ideal lensing images are generated via ray-tracing using the lens equations (Eqs.~\eqref{eq1} and \eqref{eq2}). We then incorporate instrumental effects by convolving the images with the appropriate PSF and adding noise, with both models matched to the CSST instrument specifications (Table~\ref{tab:survey_params}), to produce the final mock lensing images.

\begin{table}[h!]
  \centering
  \caption{The Key Properties of Different CSST Survey Modes: the survey area  $\Omega$, the filters used for the Main Survey Camera (MSC), the Full-Width at Half Maximum (FWHM) of the Point Spread Function (PSF), the median sky brightness in magnitude per square arcseconds, the readout noise, the total exposure time in each filter (calculated by multiplying individual exposure time with the number of exposures), the instrumental zero-point magnitude for each band, and the 5$\sigma$ limiting magnitude for a point-like source (within $R_\mathrm{EE80}$, the radius of 80\% encircled energy). The information in each brace corresponds to the respective filters.}\label{Tab1}
  \renewcommand{\arraystretch}{1.2}
  \begin{tabular}{@{}lp{0.45\textwidth}@{}}
  \toprule
  \textbf{Property} & \textbf{Values} \\ 
  \midrule
  $\Omega$ [deg$^2$] & WF: 17500, DF: 400, UDF: 9 \\
  \midrule
  Filters & $\{g, r, i, z\}$ \\
  \midrule
  PSF FWHM [$''$] & $\{0.051, 0.064, 0.076, 0.089\}$ \\
  \midrule
  Sky background & $\{22.57, 22.10, 21.87, 21.86\}$ \\
  \midrule
  Readout noise [e$^-$] & 5 \\
  \midrule
  Dark current [e$^{-}$/s] & 0.02 \\
  \midrule
  Exposure time [s] & WF: $[150 \times 2, 150 \times 2, 150 \times 2, 150 \times 2]$, \\
  & DF: $[250 \times 8, 250 \times 8, 250 \times 8, 250 \times 8]$, \\
  & UDF: $[250 \times 60, 250 \times 60, 250 \times 60, 250 \times 60]$ \\
  \midrule
  Zero point & $\{25.79, 25.60, 25.41, 24.83\}$ \\
  \midrule
  Limiting magnitude & WF: $\{26.58, 26.32, 26.03, 25.46\}$, \\
  & DF: $\{27.77, 27.50, 27.20, 26.67\}$, \\
  & UDF: $\{28.89, 28.62, 28.32, 27.78\}$ \\
  \bottomrule
  \end{tabular}
  \label{tab:survey_params}
\end{table}

Since lensing features are \xycao{usually} most prominently detected in the $g$ band \citep{cao_2024}, we select this band to simulate our mock images. We implement a Gaussian PSF with the corresponding FWHM of $g$ band as specified in Table~\ref{tab:survey_params}, , where the FWHM values are determined by fitting a Gaussian model to the simulated CSST PSF \citep{cao_2024}. For computational precision, the PSF kernel is oversampled by a factor of 16 before being subsampled back to its original size.  We adopt the same noise model as \cite{cao_2024}, with the corresponding instrumental zero-point magnitude and sky background incorporated for noise calculation.

Note that across the simulations of different survey modes, only the exposure time and the number of exposures actually differ, serving as the dominant control factors for SNR across different observational scenarios, while other conditions are kept consistent to control variables.

\begin{figure}[h!]
  \centering
  \begin{minipage}{0.33\textwidth}
      \centering
      \includegraphics[width=\linewidth]{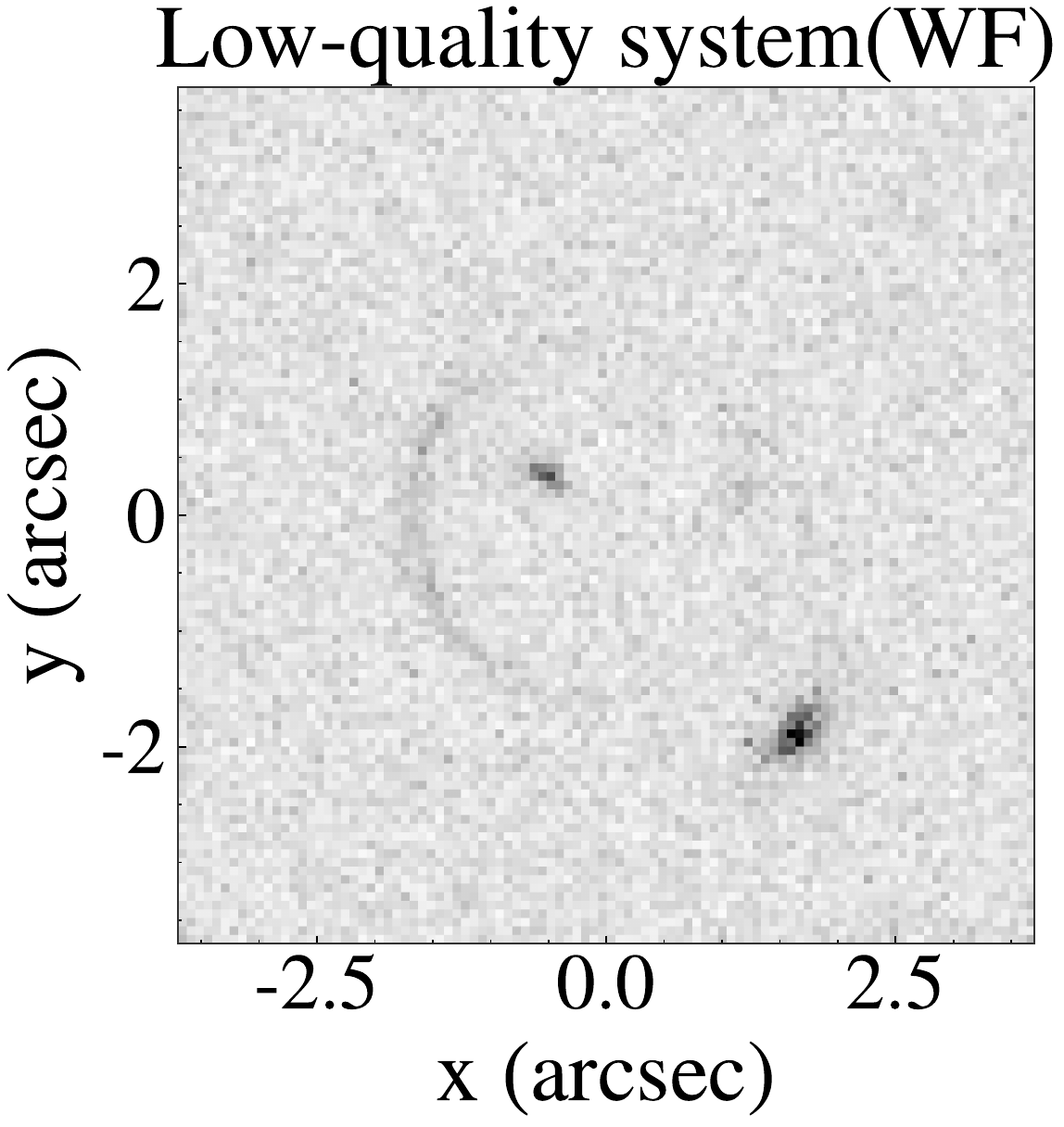}
  \end{minipage}%
  \begin{minipage}{0.34\textwidth}
      \centering
      \includegraphics[width=\linewidth]{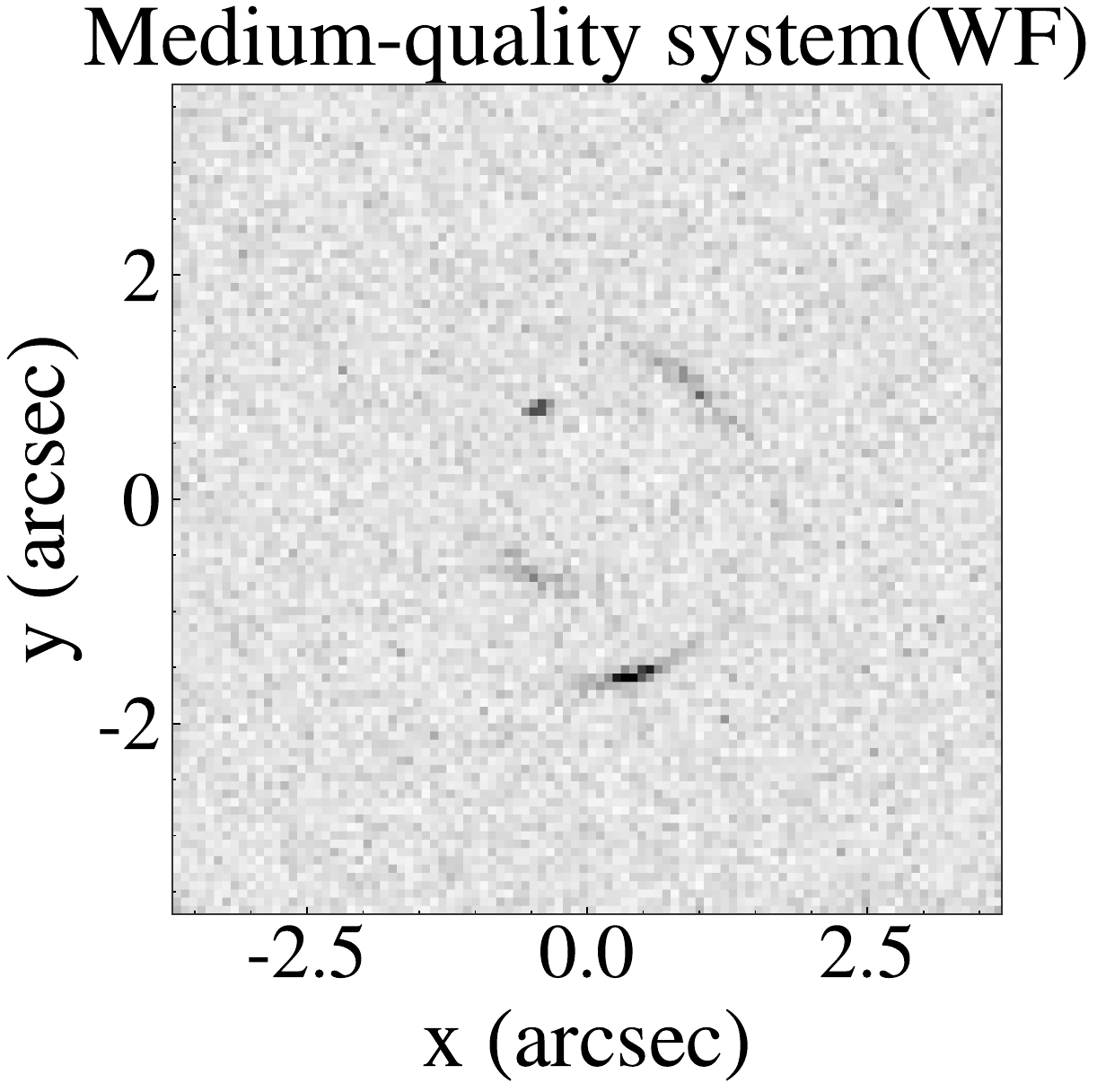}
  \end{minipage}%
  \begin{minipage}{0.33\textwidth}
      \centering
      \includegraphics[width=\linewidth]{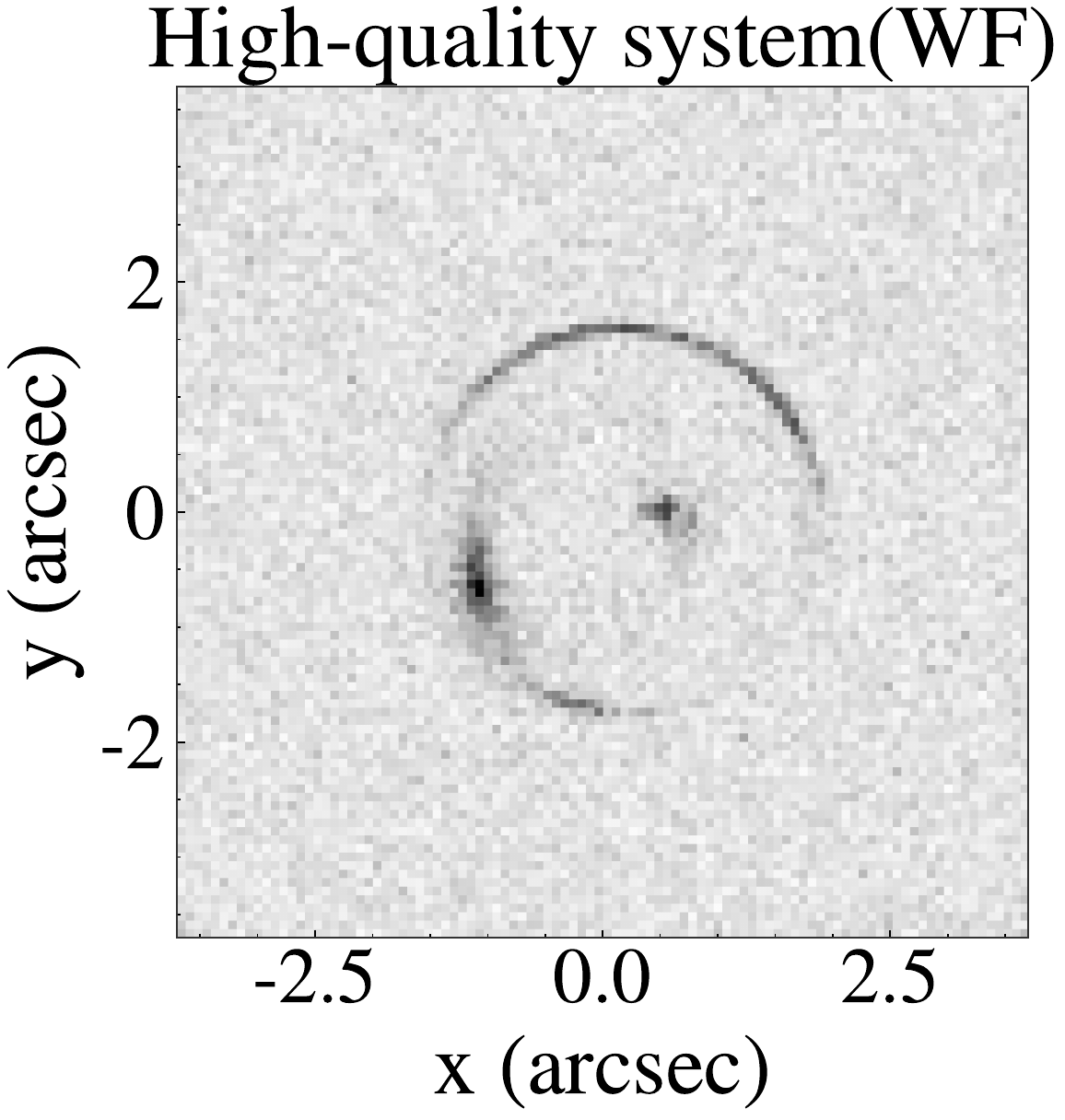}
  \end{minipage}

  \begin{minipage}{0.33\textwidth}
      \centering
      \includegraphics[width=\linewidth]{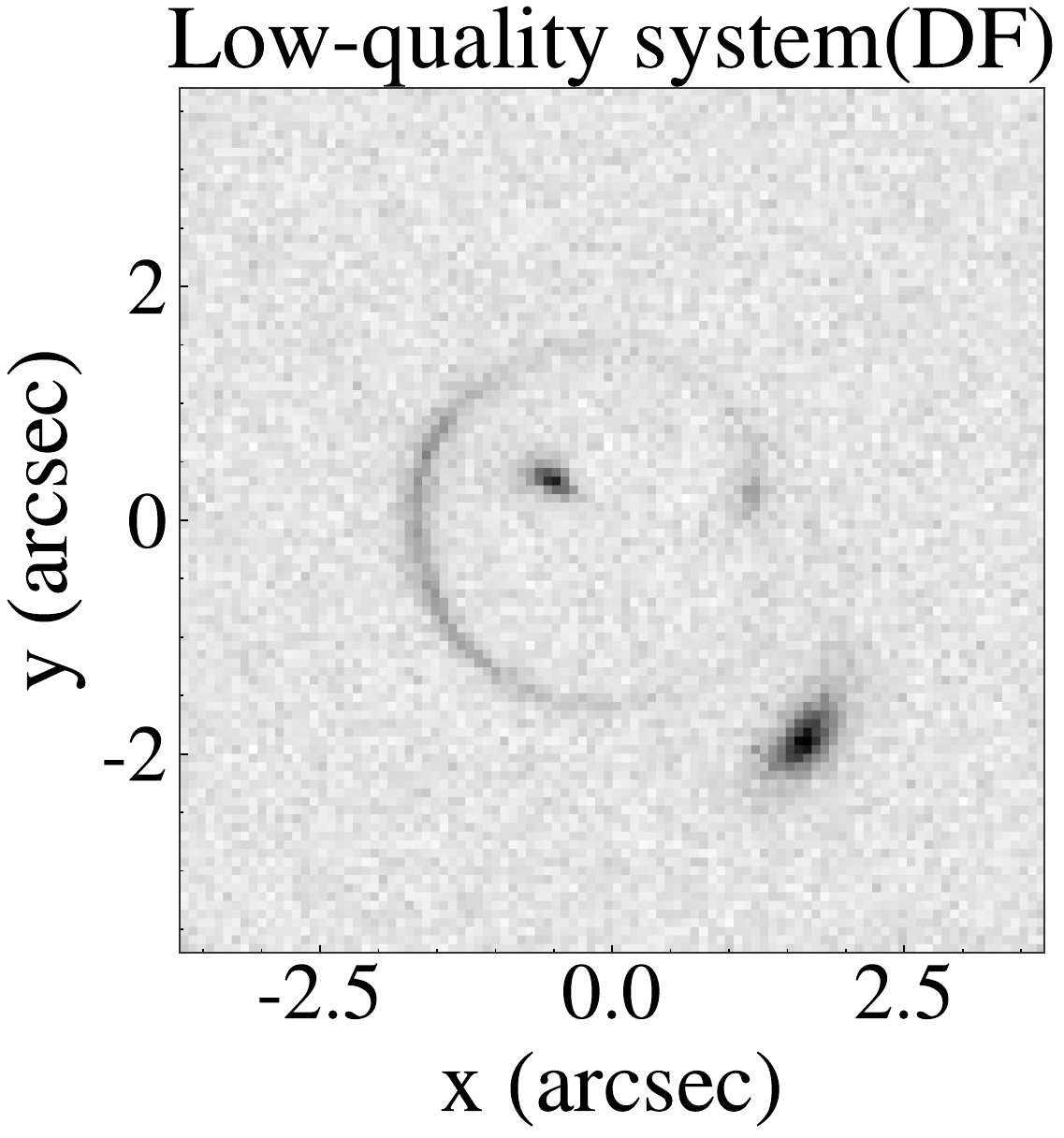}
  \end{minipage}%
  \begin{minipage}{0.34\textwidth}
      \centering
      \includegraphics[width=\linewidth]{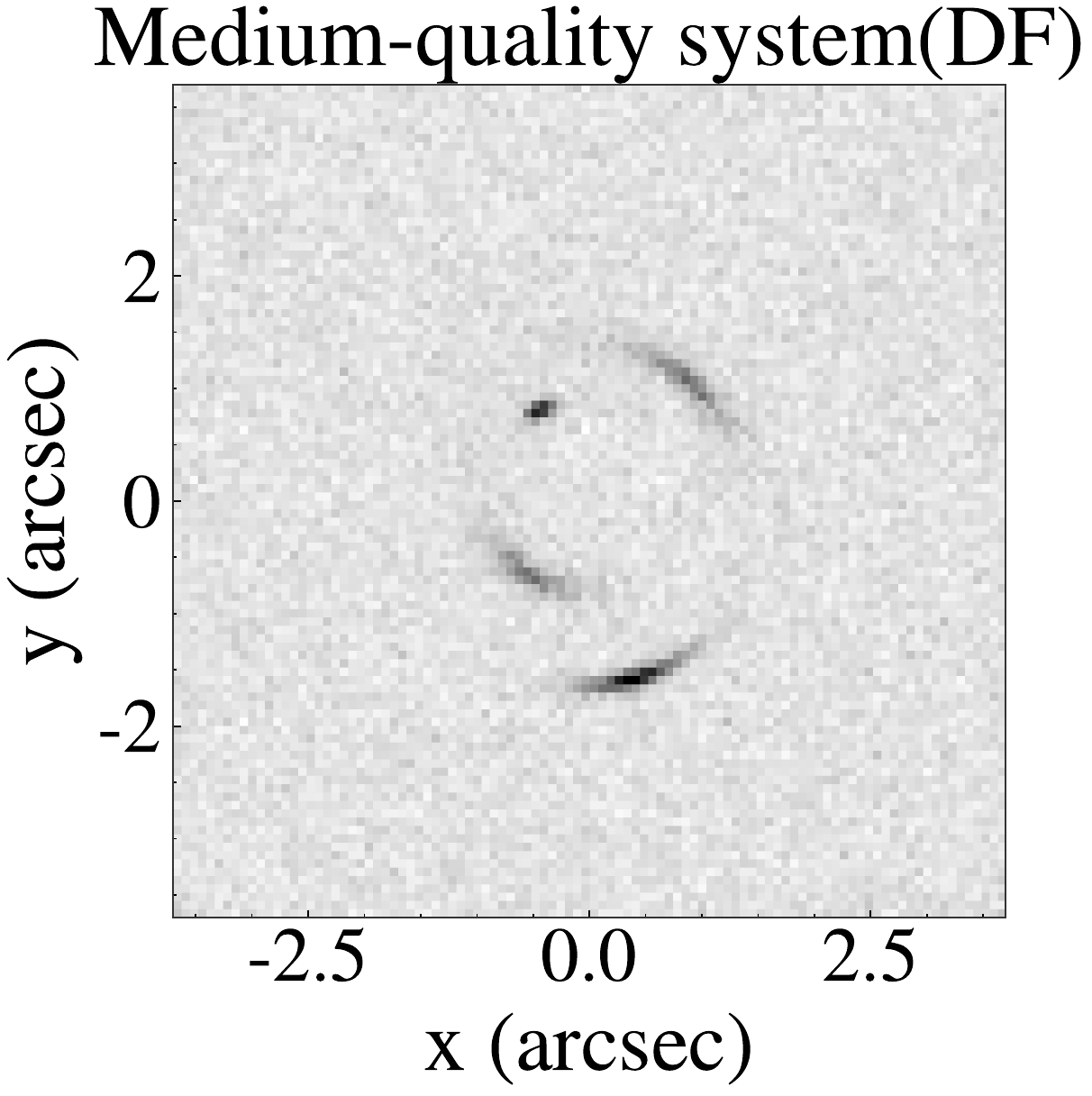}
  \end{minipage}%
  \begin{minipage}{0.33\textwidth}
      \centering
      \includegraphics[width=\linewidth]{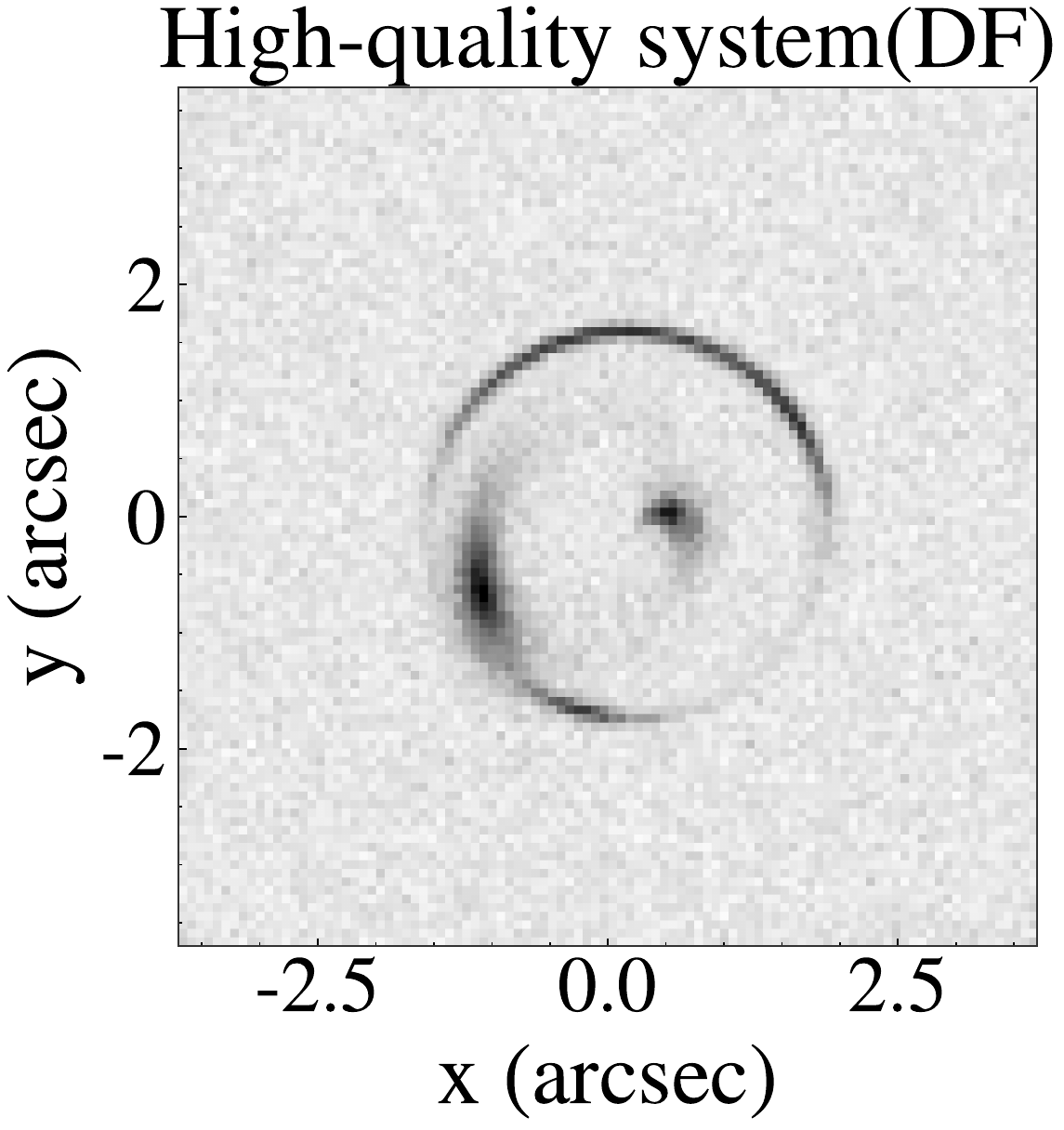}
  \end{minipage}

  \begin{minipage}{0.33\textwidth}
      \centering
      \includegraphics[width=\linewidth]{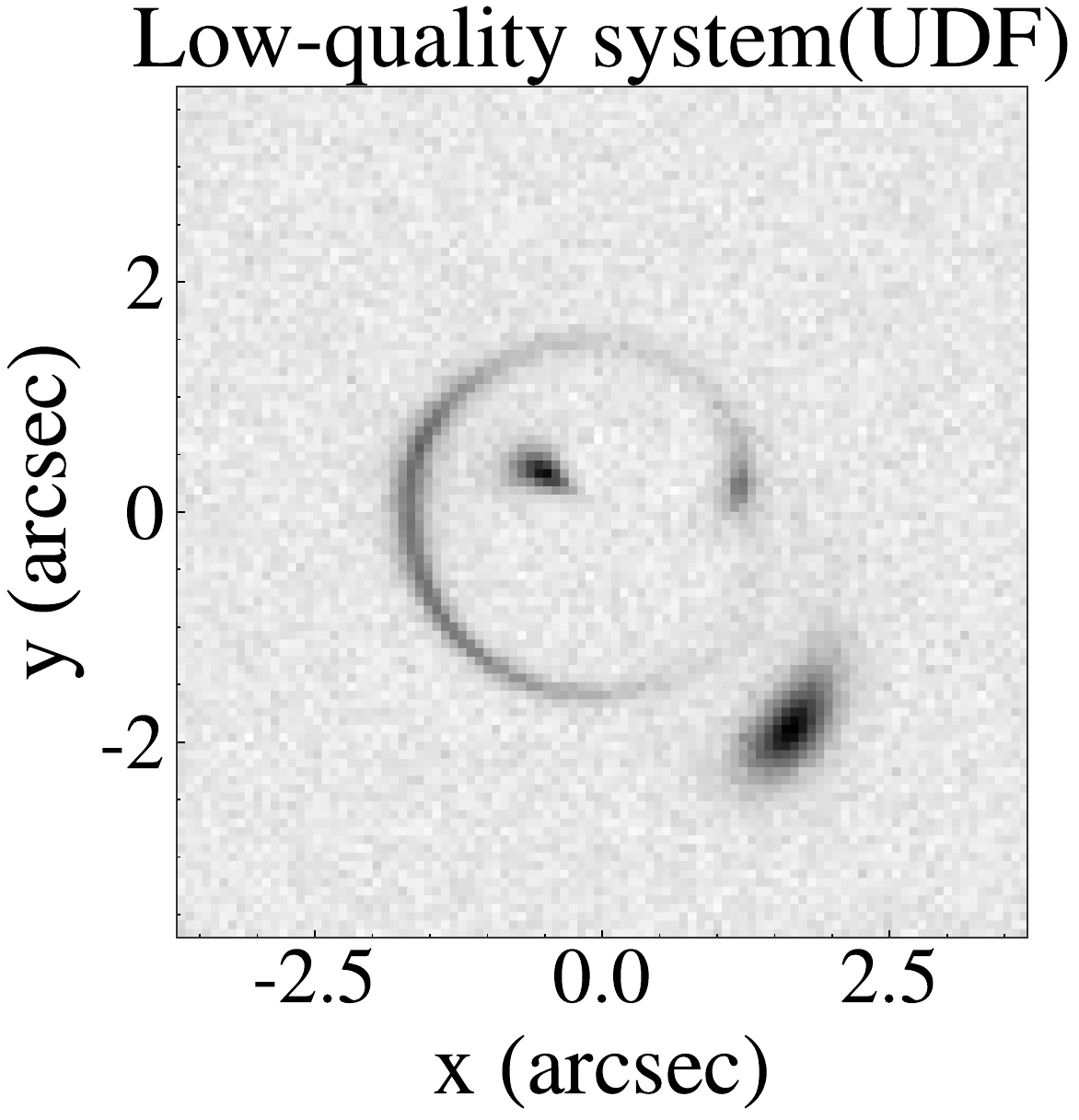}
  \end{minipage}%
  \begin{minipage}{0.34\textwidth}
      \centering
      \includegraphics[width=\linewidth]{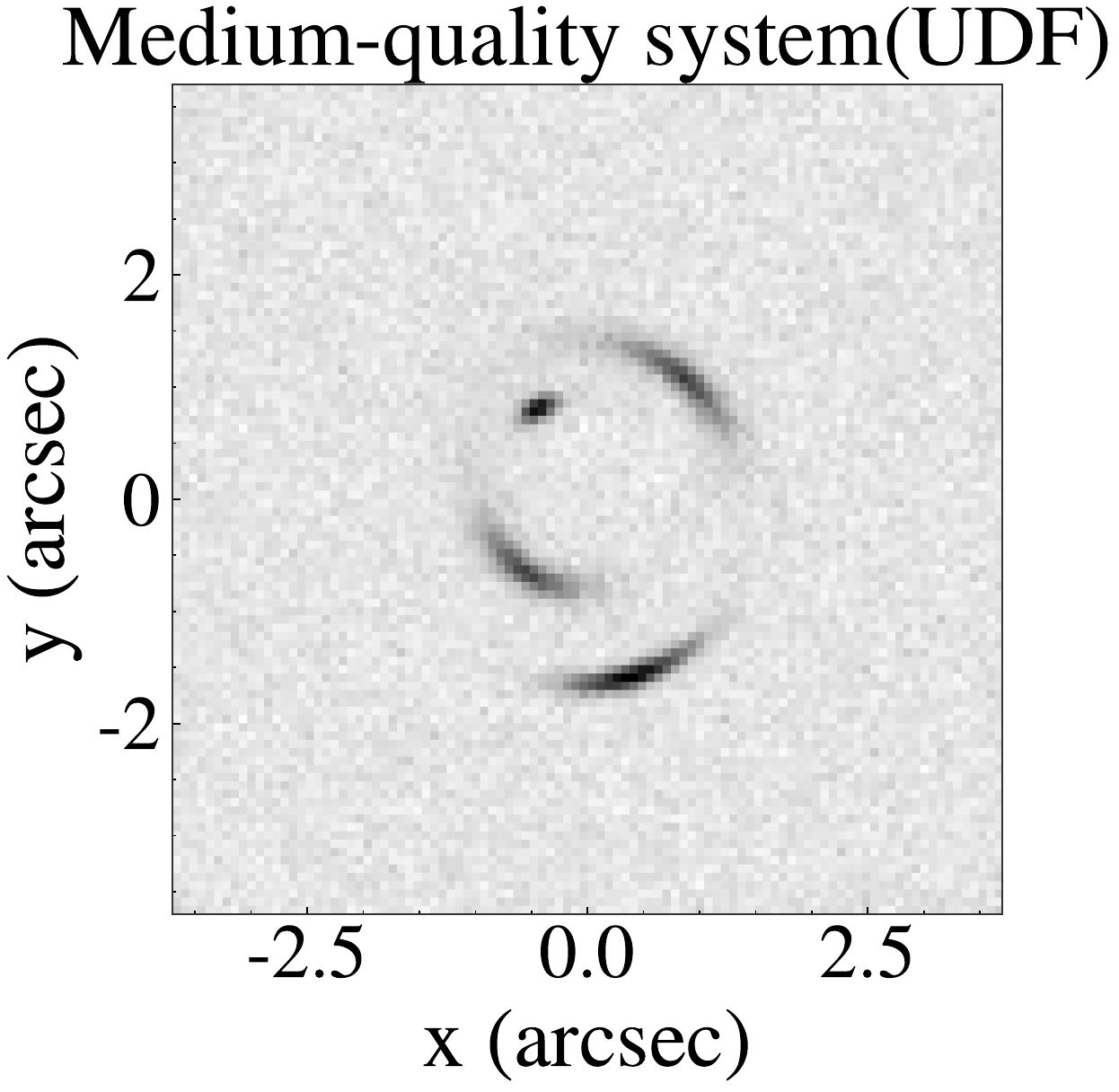}
  \end{minipage}%
  \begin{minipage}{0.33\textwidth}
      \centering
      \includegraphics[width=\linewidth]{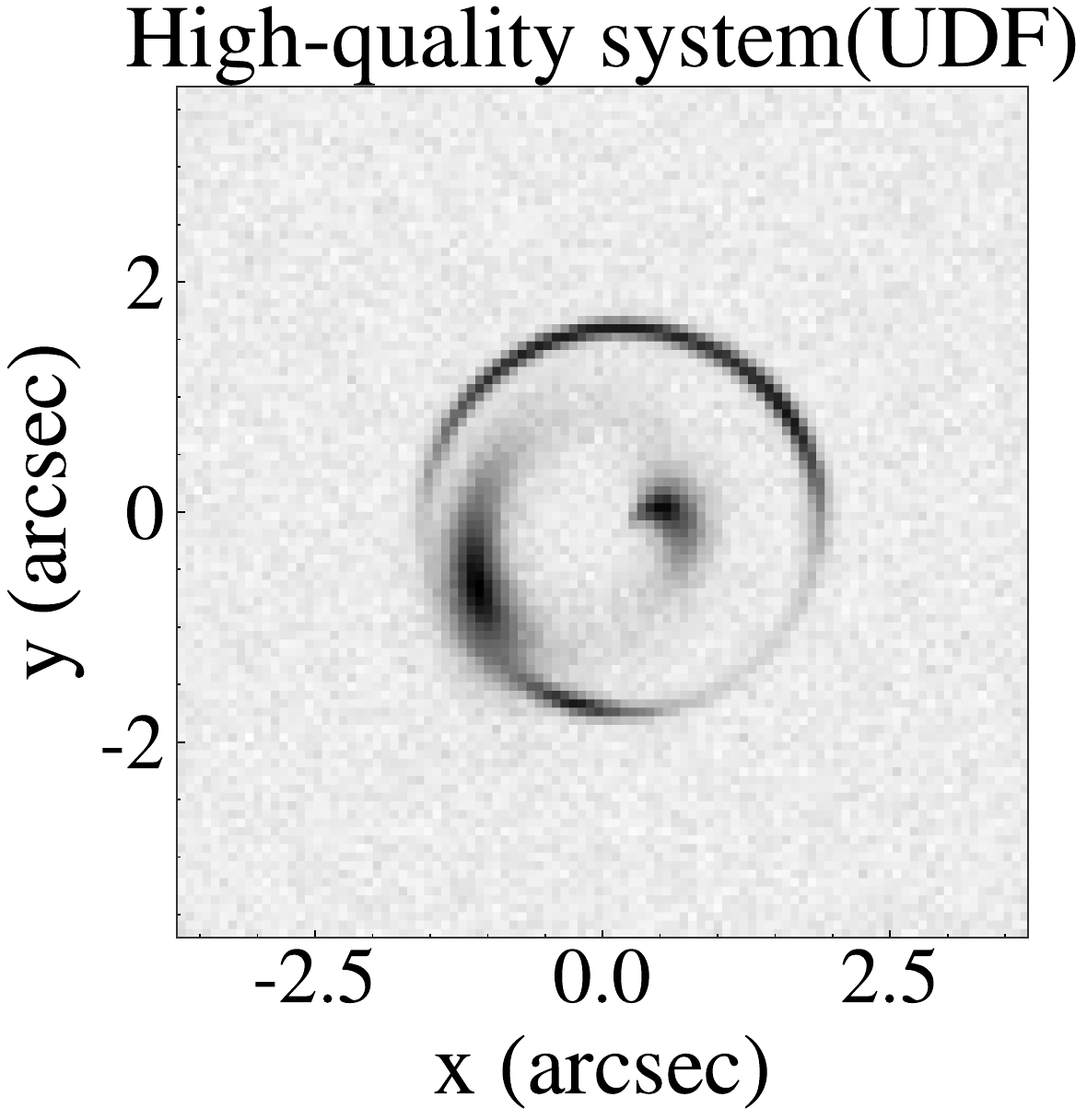}
  \end{minipage}

  \caption{The mock images of three systems from three different survey modes: WF (top row), DF (middle row), and UDF (bottom row).  Each column represents systems of different quality levels, with the left, middle, and right columns showing low-quality, medium-quality, and high-quality systems, respectively.}
  \label{figM}
\end{figure}

The generated mock images of our three selected systems across the three survey modes are presented in Fig.~\ref{figM}.

\subsection{Strong Lens Modeling}
\label{sec:Model}

Given the inter-dependencies among lens and sources in a DSPL system, modeling of all objects is conducted simultaneously. \xycao{To avoid systematic biases arising from model mismatch, we fit the mock lens systems using the same parametric forms adopted in the lensing simulations.}
Besides, while the high-precision redshift measurements are essential for effective cosmological constraints, these can be guaranteed by ground-based follow-up spectroscopic observations. Therefore, in our analysis, all redshifts are fixed to their true values.

With the redshifts fixed and assuming a flat $w$CDM universe, all remaining parameters from Table~\ref{Tab0}, along with two cosmological parameters $\Omega_{\rm m}$ and $w$ \rev{(with fiducial values $\Omega_{\rm m} = 0.30966$, $w = -1$)}, are treated as free parameters. \rev{Specifically, the centroid of the lens mass profile is treated as a free parameter to allow for potential offsets from the lens light center.} The mass and light profiles of source 1 are assumed to have the same center ($x, y$) coordinates. In total our model involves 22 free parameters, for which we assume uninformative uniform priors. For computational convenience in the model fitting process, the axis ratio $q$ and the rotation angles $\theta$ of each object are transformed into the two components of ellipticity in model fitting process, with the relation of:
\begin{equation}\label{eq8}
  \begin{aligned}
    e_1=\frac{1-q}{1+q} \cos 2 \theta, \quad \quad e_2=\frac{1-q}{1+q} \sin 2 \theta,
  \end{aligned}
\end{equation}
note that, different from \cite{collett_2014, sahu_2025}, we model our two cosmological parameters simultaneously with our 20 lensing parameters by leveraging the relation between $\beta$ and cosmological parameters to obtain cosmological constraints. To extract the effective pixels for modeling, we apply overlapping circular masks with a radius of $ 3.7''$ \rev{fixed at the center of the lens light, assuming the lens light is perfectly removed.}

Our modeling follows the parametric modeling approach of \cite{tessore_2016}, where we explore the non-linear parameter spaces for lensing parameters estimation and cosmological constraints. Define $\xi=\left\{\xi_1, \ldots, \xi_m\right\}$ as the vector of $m$ free parameters, and $d=\left\{d_1, \ldots, d_n\right\}$ as the vector of extracted pixel values in mock image. $n$ stands for the number of image pixels. Using Bayes' theorem, the posterior function can be expressed as:
\begin{equation}\label{eq9}
  \begin{aligned}
    P(\xi \mid d) \propto P(d \mid \xi) P(\xi),
  \end{aligned}
\end{equation}
assuming that the prior probabilities of the parameters are independent of one another, the prior probability $P(\xi)$ is a combination of prior probability of each parameter:
\begin{equation}\label{eq10}
  \begin{aligned}
    P(\xi) =\prod_{i=1}^m P(\xi_i),
  \end{aligned}
\end{equation}
define $L=\left\{L_1, \ldots, L_n\right\}$ to represent the corresponding vector of pixel values in model image (the blurred image without noise) generated from the corresponding parameters $\xi$, which is calculated numerically on an oversampled grid with resolution $4\times4$. Since pixel correlations can be neglected in the survey modes of the CSST Main Survey \citep{gong_2019}, the likelihood of the parameters can be expressed as:
\begin{equation}\label{eq11}
  \begin{aligned}
    P(d \mid \xi) \equiv P(d \mid L)=\prod_{i=1}^n P\left(d_i \mid L_i\right),
  \end{aligned}
\end{equation}
since the noise in our mock images is generated from a normal distribution, we have:
\begin{equation}\label{eq12}
  \begin{aligned}
    P\left(d_i \mid L_i\right)=\frac{1}{\sqrt{2 \pi s_i^2}} \exp \left\{-\frac{1}{2} \frac{\left(d_i-L_i\right)^2}{s_i^2}\right\},
  \end{aligned}
\end{equation}
where $s_i$ represents the standard deviation of each pixel.

Having established the posterior function, we implement the Markov chain Monte Carlo (MCMC) sampler $emcee$ to explore the parameter space and obtain the parameter distributions needed for cosmological constraints. For the primary focus of this research, we initialize the walkers from a uniform distribution spanning $\xi_{\text{true}} \pm 0.02\xi_{\text{true}}$ for each lensing parameter, and $\xi_{\text{true}} \pm 0.5\xi_{\text{true}}$ for each cosmological parameter.

\section{Results}
\label{sec:results}

\xycao{Fig.~\ref{fig:ca} shows the 2D distributions of the cosmological parameters inferred from MCMC analysis for system~2, whilst Fig.~\ref{fig:cb} compares the corresponding constraints obtained using lensing data observed with the UDF survey mode.} Our analysis yields the following key findings:
\begin{figure}[h]
  \centering
  \begin{subfigure}[h]{0.48\linewidth}  
      \centering
      \includegraphics[width=\linewidth]{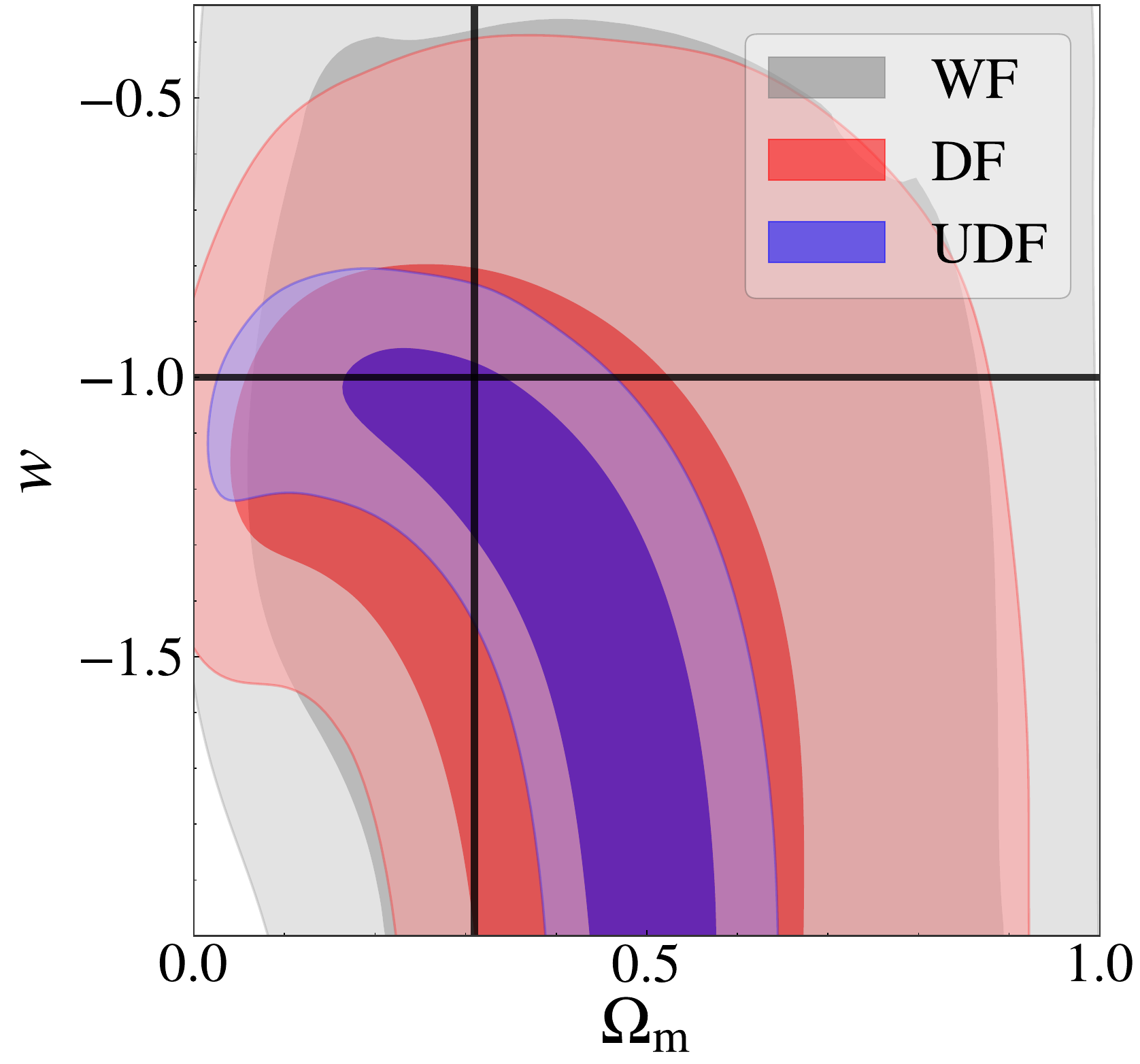}
      \caption{System 2 ($\beta^{-1}$ = 1.17)}
      \label{fig:ca}
  \end{subfigure}
  \hspace{0.1cm}  
  \begin{subfigure}[h]{0.48\linewidth}  
      \centering
      \includegraphics[width=\linewidth]{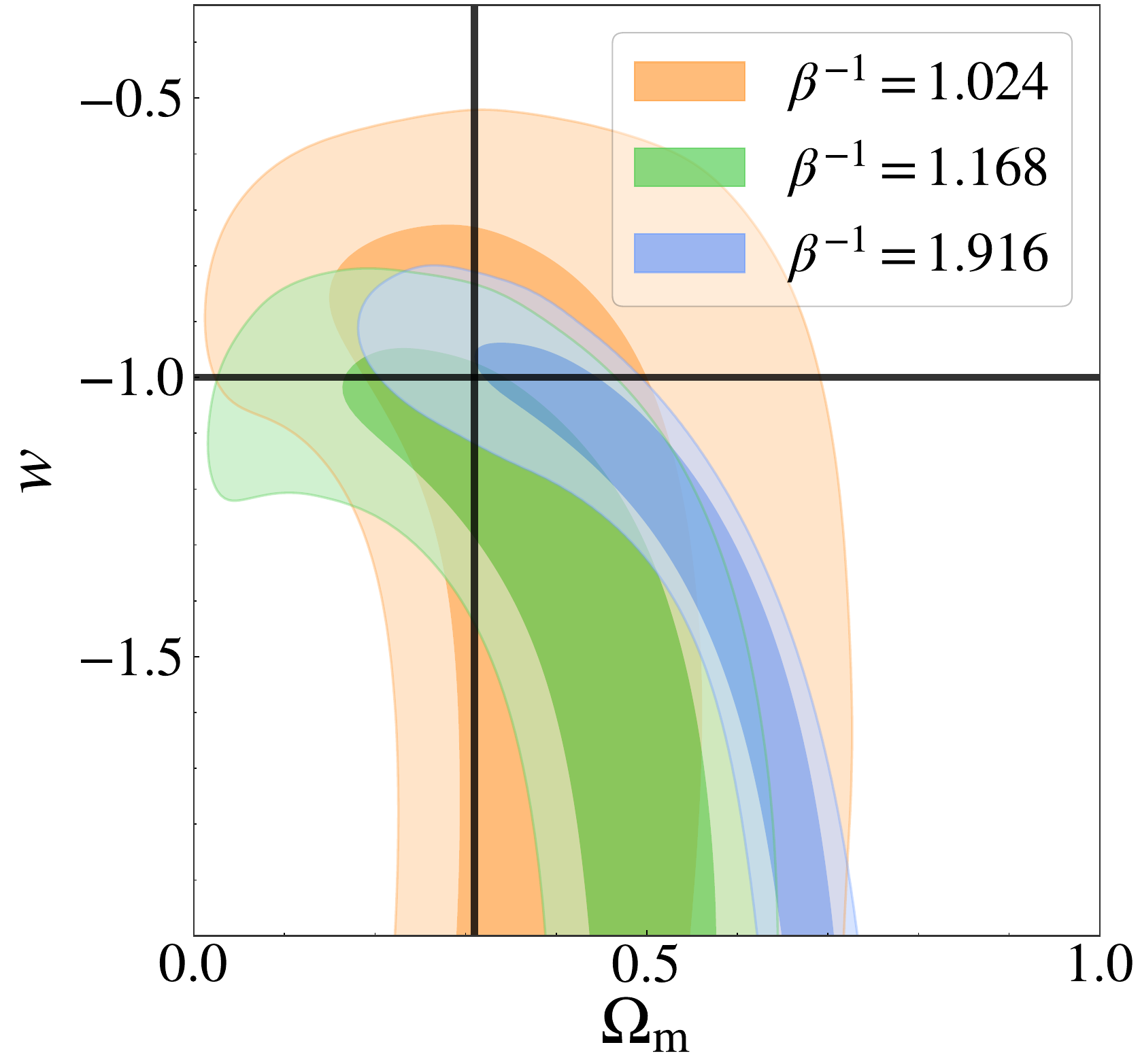}
      \caption{Three selected systems under UDF}
      \label{fig:cb}
  \end{subfigure}
  \hspace{0.1cm}  
  \caption{Cosmological constraints of: (a) System 2 across three survey modes (wide field (WF), deep field (DF), and ultra deep field (UDF)). (b) three systems under UDF. \rev{The inner and outer shaded regions represent the 68\% and 95\% confidence regions.}}
  \label{FigC}
\end{figure}

Our results demonstrate that the constraint precision on cosmology by the DSPL system increases from wide-field, deep-field, to ultra-deep-field CSST observations. The System 2, as a typical DSPL system with two distinct lensing arcs and a moderate $\beta$ value, provides the following cosmological constraints: (\(w = -1.28_{-1.00}^{+0.64}\), \(\Omega_{\rm m} = 0.50_{-0.32}^{+0.28}\)), (\(w = -1.51_{-0.54}^{+0.54}\), \(\Omega_{\rm m} = 0.44_{-0.20}^{+0.20}\)), and (\(w = -1.59_{-0.32}^{+0.63}\), \(\Omega_{\rm m} = 0.42_{-0.06}^{+0.15}\)) for WF, DF, and UDF. The corresponding constraints on $\beta^{-1}$ are \(\beta^{-1} = 1.170_{-0.0045}^{+0.0042}\), \(\beta^{-1} = 1.168_{-0.0030}^{+0.0031}\), and \(\beta^{-1} = 1.168_{-0.0011}^{+0.0011}\).

Furthermore, our analysis reveals that under identical survey characteristics, different double-ring features provide varying constraints, with DSPL systems having larger $\beta^{-1}$ values yielding significantly tighter constraints. \xycao{This trend has also been discussed by \citet{collett_2012} and \citet{linder_2016}}. This underscores the critical importance of system selection when choosing a single DSPL for cosmological constraints. As shown in Fig.~\ref{fig:cb}, for systems with $\beta^{-1} = 1.92$, $\beta^{-1} = 1.17$ and $\beta^{-1} = 1.02$, the cosmological constraints under UDF are (\(w = -1.51_{-0.26}^{+0.59}\), \(\Omega_{\rm m} = 0.55_{-0.06}^{+0.16}\)), (\(w = -1.59_{-0.32}^{+0.63}\), \(\Omega_{\rm m} = 0.42_{-0.06}^{+0.15}\)) and (\(w = -1.52_{-0.44}^{+0.74}\), \(\Omega_{\rm m} = 0.41_{-0.14}^{+0.12}\)), respectively. The corresponding constraints on $\beta^{-1}$ are \(\beta^{-1} = 1.916_{-0.0018}^{+0.0018}\), \(\beta^{-1} = 1.168_{-0.0011}^{+0.0011}\), and \(\beta^{-1} = 1.024_{-0.0009}^{+0.0009}\). \xycao{Since} $\beta^{-1}$ represents the ratio of the two Einstein radii in DSPL under simplified assumptions, it can be approximately estimated from observational images during system selection.

\rev{Compared with previous observational results, our analysis demonstrates the potential for tighter constraints on cosmological parameters using DSPLs from the CSST survey.} Assuming a flat $\Lambda$CDM cosmology, \rev{we infer \(\Omega_{\rm m} = 0.35_{-0.05}^{+0.05}\) for high-quality system (System 1) under UDF (\(\beta^{-1} = 1.916_{-0.0018}^{+0.0018}\))}, which is approximately six times tighter than the DSPL example SDSSJ0946+1006 \rev{(the ``Jackpot'' case, \(\beta^{-1} = 1.405_{-0.016}^{+0.014}\))} reported by \citet{collett_2014} and the second DSPL ever used to measure cosmological parameters \rev{(AGEL150745+052256, \(\beta^{-1} = 1.049_{-0.009}^{+0.011}\))} reported by \citet{sahu_2025}, and five times tighter than the DSPL AGEL035346-170639 \rev{(\(\beta^{-1} = 1.076_{-0.004}^{+0.010}\))} \citep{Bowden_2025}. 

It is important to emphasize that these results represent forecasts under idealized scenarios. Additional systematic errors must be taken into consideration in actual observations. Beyond the inherent degeneracies and systematic errors introduced by models, redshift uncertainties have non-negligible effects on cosmological constraints, particularly when using a single DSPL system for cosmological constraints.

\section{Discussion and Conclusions}
\label{sec:conclusion}

In this work, we explored the \xycao{capabilities} and robustness of DSPLs in constraining cosmological parameters in the era of CSST by using mock data. The outcomes demonstrate that the precision of cosmological constraints improves systematically across WF, DF, and UDF, with DSPL systems of larger $\beta^{-1}$ value providing substantially tighter constraints. Therefore, during the early stages of CSST operation, before a sufficient DSPL sample for hierarchical Bayesian analysis is available, the DSPL systems with larger $\beta^{-1}$ value in UDF are recommended prioritizing for cosmological research. 

Specifically, a typical DSPL system with two distinct lensing arcs and a moderate $\beta^{-1}$ value yields the following cosmological constraints across WF, DF, and UDF: (\(w = -1.28_{-1.00}^{+0.64}\), \(\Omega_{\rm m} = 0.50_{-0.32}^{+0.28}\)), (\(w = -1.51_{-0.54}^{+0.54}\), \(\Omega_{\rm m} = 0.44_{-0.20}^{+0.20}\)), and (\(w = -1.59_{-0.32}^{+0.63}\), \(\Omega_{\rm m} = 0.42_{-0.06}^{+0.15}\)); and for systems with $\beta^{-1} = 1.92$, $\beta^{-1} = 1.17$ and $\beta^{-1} = 1.02$ (which represents varying ratio of the two Einstein radii), the cosmological constraints under UDF are (\(w = -1.51_{-0.26}^{+0.59}\), \(\Omega_{\rm m} = 0.55_{-0.06}^{+0.16}\)), (\(w = -1.59_{-0.32}^{+0.63}\), \(\Omega_{\rm m} = 0.42_{-0.06}^{+0.15}\)) and (\(w = -1.52_{-0.44}^{+0.74}\), \(\Omega_{\rm m} = 0.41_{-0.14}^{+0.12}\)), respectively.

However, it is worth noting that, due to the limited survey area of UDF ($\sim 9\,\text{deg}^2$) \citep{gong1_2025}, there might be a limited number of DSPL found in the UDF field according to the probability of DSPL systems given by \cite{cao_2024}. Therefore, we recommend that the CSST-DF field is optimal for finding high-quality DSPLs and utilize them to constrain cosmological parameters. Besides, significant challenges remain in achieving precise modeling in real observational scenarios, with relevant research in this area still notably insufficient. Our work serves primarily as an optimistic forecast under idealized conditions. In future work, based on state-of-the-art observations and simulations, we will conduct more sophisticated analyses with more realistic facts being taken into account.

Nonetheless, an individual DSPL system remains a \rev{valuable} cosmological probe with a lot of potential applications in the next-generation survey era. Beyond CSST, the Euclid, Roman, and JWST telescopes all possess capabilities to utilize DSPL for cosmological constraints. Joint observational campaigns across these platforms will further improve measurement precision. Additionally, synergistic observations combining JWST with ground-based extremely large telescopes can significantly enhance the cosmological constraints derived from DSPL systems \citep{sharma_2023}. 

Besides, as larger DSPL sample accumulate in the future, statistical methods can be applied to Wide Field and Deep Field observations to enhance constraint precision. Statistical methods such as hierarchical Bayesian analysis can substantially improve DSPL constraints in Wide Field and Deep Field surveys. Given the potential number of DSPL systems that CSST could detect across different survey modes, we recommend employing statistical techniques with multiple DSPL systems to obtain more robust cosmological constraints (further discussion on this approach will be presented in subsequent work). \rev{ In future work involving hierarchical Bayesian analysis of large DSPL samples, combining these constraints with CMB priors will be a highly effective approach to break degeneracies and achieve high-precision cosmology.}

Furthermore, large sample sizes lead to opportunities for discovering rare DSPL systems with additional information beyond images and spectrum only. A noteworthy example is the recently confirmed first Einstein zig-zag lens, which features multiple lensed sources resulting in two separate Einstein radii of different sizes, including a variable quasar. As the first DSPL system with time-delay information, it offers an extremely rare opportunity to combine two major lensing cosmological probes to better constrain both mass profiles of the lenses and cosmological parameters \citep{dux_2024}. 

To summarize, DSPL systems are valuable for astrophysics and cosmology. In particular, the upcoming large-scale surveys, like CSST and Roman, bring a large sample of DSPLs, enabling more accurate and precise constraints on the properties of dark matter and dark energy. However, the selection criterion is critical for such studies. In this study, focusing on CSST, we present that the constraints are sensitive to the signal noise ratio of the observations and the ratio of the two Einstein radii, which suggests that deep observations are required and the DSPL systems with larger ratio of the two Einstein radii are optimal, if one adopt individual DSPL systems for astrophysical and cosmological studies. Yet, it is still an open question that the best strategy for combining a large sample of DSPL systems to implement joint constraints on the parameters of the models of dark matter and cosmology, and we will conduct thorough investigation on it in future work.     

\begin{acknowledgements}
BCW and YG acknowledge the support from the CAS Project for Young Scientists in Basic Research (No. YSBR-092), National Key R\&D Program of China grant Nos. 2022YFF0503404 and 2020SKA0110402. XYC acknowledges the support of the National Natural Science Foundation of China (NSFC) under No.12303006. This work is also supported by science research grants from the China Manned Space Project with grant Nos. CMS-CSST-2025-A02, CMS-CSST-2021-B01, and CMS-CSST-2021-A01.
\end{acknowledgements}



\bibliographystyle{raa}
\bibliography{bibtex}

\label{lastpage}

\end{document}